\documentstyle[twocolumn,prl,aps]{revtex}
\begin{document}
\draft

\title{ Quantized vortices in mixed $^3$He-$^4$He drops
}
\author{Ricardo Mayol, Mart\'{\i} Pi, and Manuel Barranco}

\address{
Departament ECM,  Facultat de F\'{\i}sica,
Universitat de Barcelona, E-08028 Barcelona, Spain}

\author{Franco Dalfovo}

\address{
Dipartimento di Matematica e Fisica, Universit\`a Cattolica,
and Istituto Nazionale
\\
per la Fisica della Materia, Unit\`a di Brescia,
Via Musei 41, I-25121 Brescia, Italy}

\date{\today}

\maketitle

\begin{abstract}
Using density functional theory, we investigate the structure of  
mixed $^3$He$_{N_3}$-$^4$He$_{N_4}$ droplets with an embedded 
impurity (Xe atom or HCN molecule) which pins a quantized vortex 
line.  We find that the dopant+vortex+$^4$He$_{N_4}$ complex,
 which in a previous work [F. Dalfovo {\it et al.}, Phys. Rev.
Lett. {\bf 85}, 1028 (2000)] was found to be energetically stable
below a critical size $N_{\rm cr}$,
is robust against the addition
of $^3$He. While $^3$He atoms are distributed along the vortex line 
and on the surface of the $^4$He  drop, the impurity is mostly 
coated by $^4$He atoms. Results for $N_4=500$ and a number of 
$^3$He atoms ranging from 0 to 100 are presented, and the binding 
energy of the dopant to the vortex line is determined.

\end{abstract}


Helium nanodroplets have recently attracted a considerable
interest. A major reason is the possibility of using them
as an inert, ultracold matrix for molecular spectroscopy studies
\cite{Goy92}. They also allow one to
address superfluid phenomena at a microscopic scale\cite{Toe01},
and constitute an ideal testing ground for quantum many-body theories.
An interesting perspective in this direction is the investigation of 
quantized vortices in finite systems.  Key experiments
have been carried out in the last two years on vortices in
Bose-Einstein condensed gases of rubidium and sodium atoms confined
in magnetic traps \cite{bec}, where vortical states are created by
acting with external perturbations in different ways. These states 
turn out to be more robust than expected on the basis of qualitative 
arguments. In principle,  analogous vortical configurations are also 
possible in superfluid $^4$He droplets, where the external 
perturbation could be a moving and/or rotating impurity. Although the
presence of vortices in superfluid $^4$He drops is not energetically
favorable\cite{Bau95}, we have argued that they can be stabilized by 
molecules hosted in the bulk of the drop\cite{Dal00}, and that their 
existence could be inferred from the changes they induce in the 
molecular spectrum\cite{Nau99}.

The aim of this work is to extend our previous analysis \cite{Dal00} 
to the case of mixed $^3$He-$^4$He droplets. The addition of $^3$He 
atoms to doped $^4$He droplets has significant consequences in current 
experiments, since it lowers  the temperature of the droplet from 
about $0.4$ K to $0.15$ K, and their presence can be used as another 
source of information for characterizing the interaction of
the dopant with the superfluid environment \cite{Gre98}. In
this context, an accurate description of the first solvation
layers of $^4$He and/or $^3$He around the impurity is important.
The presence of $^3$He in a droplet hosting a quantized vortex
would display several interesting features, since: i) $^3$He atoms 
behave as a normal component in the superfluid, providing a friction 
mechanism for the motion of vortex lines; ii) $^3$He atoms occupy 
surface states, known as Andreev states, which are energetically 
favored by the lighter mass and the larger zero point motion of 
$^3$He compared to $^4$He; iii) some $^3$He atoms will be attached to
the vortex core, where they are expected to have a binding energy of the
order of 2-3 K \cite{Donnelly}. We investigate the effects (ii) and
(iii) by using a density functional method to calculate the
structure and the energetics of these systems, also including
dopant atoms and molecules. In the calculations we use Xe or HCN
embedded in droplets with $N_4=500$ atoms of $^4$He and a number
of $^3$He atoms $N_3$ ranging from $0$ to $100$.

Our starting point is a density functional previously developed
for mixed $^3$He-$^4$He systems, which allows one to write
the energy of the mixture as $E = \int d{\bf r}\, {\cal H} 
[\rho_3({\bf r}),\rho_4({\bf r})]$, where $\rho_3({\bf r})$ 
[$\rho_4({\bf r})]$ is the $^3$He [$^4$He] particle density (see 
\cite{Bar97,Pi99} and references therein). To keep the already 
cumbersome calculations at an affordable level, here we use a 
slightly simplified version of the same functional; namely, we take 
the core of the screened Lennard-Jones He-He potentials as in the 
original Orsay-Paris functional \cite{Dup90}, and drop the 
gradient-gradient term which appears in the Orsay-Trento
functional\cite{Dal95} for $^4$He. We have checked that these changes
have negligible effects on the relevant results, 
while drastically reduce the
numerical effort. As discussed in \cite{Bar97,Pi99}, the density 
functional contains a set of parameters which are fixed to reproduce  
static properties of pure and mixed He systems at zero temperature, 
like the equation of state, surface tension of the different 
interfaces, the osmotic pressure and maximum solubility of $^3$He 
into $^4$He\cite{T=0}.

As in \cite{Dal00}, the vortex line is included through the
Feynman-Onsager  {\it ansatz}, i.e., by  adding an extra
centrifugal energy associated with the velocity field of $^4$He, 
which is singular on the vortex axis, thus forcing its density to 
vanish.  For doped droplets, one has to include the 
helium-impurity interaction, which acts as an external potential 
in which the helium density adjusts to minimize the energy. The 
potential for Xe has been taken from \cite{Tan86}, and that of 
HCN from \cite{Atk96}. Combining all terms, the total energy
can be written in the form
\begin{eqnarray}
E&=&\int d{\bf r}\, \left\{ {\cal H} [\rho_3({\bf r}),\rho_4({\bf r})] +
\frac{\hbar^2}{ 2 m_4 r_\perp^2} \rho_4({\bf r}) \right. \\
\nonumber
&+& \left. V_I({\bf r}) [\rho_3({\bf r})+\rho_4({\bf r})]\right\} \;\, ,
\label{eq1}
\end{eqnarray}
where  $V_I$ is the helium-impurity potential and $r_\perp$
is the distance from the vortex axis. The  energy minimization is 
performed in axial symmetry by mapping the densities on a spatial mesh,
putting the vortex line along the $z$-axis and the dopant in the 
center, at ${\bf r}=0$.

First we consider two rather simple configurations: i) a mixed
droplet with a dopant, but no vortex; ii) a mixed droplet with
a vortex, but no dopant. In the former case we just re-obtain
the results of \cite{Bar97,Pi99}, namely that the amount of $^3$He
atoms in the bulk of the drop is negligible, and the dopant is
coated by $^4$He.  The  structure of the drop is `onion like',
with $^3$He distributed in the outer shell and
$^4$He inside, surrounding the embedded impurity. In contrast, in the
latter case $^3$He atoms can funnel through the vortex dimple
created at the $^3$He-$^4$He interface and eventually
the vortex core is filled with $^3$He.

An example of a mixed droplet with a vortex is shown in Fig.~\ref{fig1}.
The $^4$He and $^3$He density profiles in the radial direction, at $z=0$,
are shown in Fig.~\ref{fig2} for different values of $N_3$.  It
can be seen that $^3$He is filling the vortex core even for $N_3=20$,
and that the dependence of the central density on $N_3$ is weak, as
expected for a close-packed linear chain of atoms. The remaining  
$^3$He atoms occupy the available surface states, whereas $^4$He 
stays in the bulk.
A comparison with the case of pure $^4$He droplets (solid line in
Fig.~\ref{fig2}) shows that $^3$He  atoms  push the superfluid
$^4$He component away from the vortex axis, thus lowering the kinetic
energy associated with the vortex flow \cite{Jez97,note1}.

Placing a dopant in the center of the droplet significantly distorts
the $^3$He and $^4$He densities. In Fig.~\ref{fig3} we show the same 
densities as in Fig.~\ref{fig1}, but with an embedded HCN molecule. 
The strong helium-impurity attraction,
which is the same for $^3$He and $^4$He, favors the formation of
a layered structure of $^4$He atoms near the molecule, since they
have a smaller zero point motion than $^3$He atoms and can be
localized more easily in the local minima of the potential.
In the vicinity of the dopant, $^3$He atoms only remain at the pinning
points on the vortex axis, where $^4$He is excluded from by the high
kinetic energy of the vortex  flow.  The net effect is that the
structure of the complex dopant+vortex near the dopant is very 
similar to that of pure $^4$He droplets. It is worth to see that the
dopant also produces a modulation of the $^3$He density along the 
vortex line.

Let us denote with subscripts $X$ and $V$ the energies, $E$, of mixed 
droplets doped with an impurity $X$ and/or containing a vortex line. 
The energetics of these droplets can be studied by 
introducing, for a fixed value of $N_4$, the following functions of 
$N_3$ \cite{Dal00}:
\begin{eqnarray}
\Delta E_{\rm V} (N_3) &=& E_{\rm V}(N_3) -E(N_3)
\label{eq2} \\
\Delta E_{\rm V}^{\rm X} (N_3)  &=& E_{\rm X+V} (N_3) - E_{\rm X} (N_3)
\label{eq3}\\
S_{\rm X} (N_3) &=&  E_{\rm X} (N_3) -E (N_3)
\label{eq4}\\
S_{\rm X+V}(N_3) &=&  E_{\rm X+V} (N_3) -E (N_3)
\label{eq5}\\
\delta_{\rm X} (N_3) &=&
\Delta E_{\rm V}^{\rm X}(N_3) - \Delta E_{\rm V}(N_3) \; .
\label{eq6}
\end{eqnarray}
The quantities $\Delta E_{\rm V}$ and $\Delta E_{\rm V}^{\rm X}$
correspond to the energy associated with the vortex flow in
a droplet without dopant and with dopant $X$, respectively.
The quantity $S_X$ is the solvation energy of impurity $X$ in
the mixed cluster, while $S_{\rm X+V}$ is the solvation energy of
the dopant+vortex complex. When the quantity $\delta_{\rm X}$ is
negative, its absolute value represents the binding energy of the
dopant to the vortex in the mixed cluster.

In Ref.~\cite{Dal00} we have studied the above energies in the
$N_3=0$ case, finding that $S_{\rm X+V} (N_3=0)$ is negative and hence 
the $^4$He+X+vortex complex is stable for values of $N_4$ smaller 
than a critical number, $N_{cr}$, of the order of $8000$ for both Xe 
and HCN. The effect of a non-zero $N_3$ value comes from a delicate
interplay between different energy contributions, which are sensitive
to the distribution of $^3$He atoms in the vortex and near the 
dopant. The crucial question is whether $^3$He may depin the impurity 
from the vortex core, i.e., $\delta_{\rm X}(N_3)$ becomes positive 
for a certain $N_3$.

The relevant energies are plotted in Fig.~\ref{fig4}.
The top panel
shows how the vortex energy decreases with $N_3$ in droplets
without dopant, with a Xe atom, and with a HCN
molecule. This behavior is consistent with a reduction
of the kinetic energy of the vortex flow when normal $^3$He 
atoms displace superfluid $^4$He atoms away from the vortex core.
The two middle panels show the solvation energy of the impurity
and of the dopant+vortex
complex; both are  negative for these droplets.
The main result of
this analysis is shown in the bottom panel, where one may see the
effect of $^3$He on the binding energy of the dopant to the vortex.
The binding energy $|\delta_X|$ of the dopant decreases when  $N_3$
increases, but the dopant is still pinned to the vortex. The
initial slope of the curve is steeper than for large $N_3$. This
is consistent with  the first $^3$He atoms occupying
states along the vortex line close to the dopant, thus affecting
$\delta_X$ in a more significant way. In the intermediate region
for $N_3=20$-$40$, the binding energy exhibits a plateau. In this
range, the presence of the dopant is expected to affect the
distribution of $^3$He atoms both in the vortex line and at the
surface. One can see in Fig.~\ref{fig2} that in the same range of
$N_3$ values the mixed droplet starts to develop a $^3$He `skin', i.e.,
an outer shell where the $^3$He density is larger than the $^4$He
density. Whereas its effect is imperceptible at the scale of the
energies defined in Eqs. \ref{eq2}-\ref{eq5}, it shows up in
$\delta_X$, which is around two orders of magnitude smaller
\cite{note2}.

We conclude that the dopant+vortex+$^4$He$_{N_4}$ complex
is robust against the addition of
moderate amounts of $^3$He atoms. This may offer some 
experimental advantages. On the one hand, mixed droplets reach lower
temperatures than pure $^4$He droplets; on the other hand,
adding a variable amount of the normal component will result in a
variable damping for the vortex motion, without loosing
the  characteristics that make $^4$He drops appealing for 
molecular spectroscopy, since the dopant environment essentially 
consists of $^4$He atoms as in pure drops.

We thank Francesco Ancilotto for an useful suggestion that allowed 
us to improve the perfomance of the original numerical routines.
This work has been performed under grants No. PB98-1247 from DGESIC, 
Spain, and No. 2000SGR-00024 from Generalitat of Catalunya, and with
the support of the Ministero dell'Universit\`a e della Ricerca 
Scientifica.

\begin{figure}
\caption{ Density distributions of $^3$He (top) and $^4$He (bottom)
in the $xz$ plane for the $^4$He$_{500}$+$^3$He$_{100}$ droplet 
hosting a vortex line along the $z$ axis.  Lengths are in units of 
{\rm \AA}. Darker regions are high density regions.}
\label{fig1}
\end{figure}

\begin{figure}
\caption{ Density profiles of $^4$He and $^3$He in the radial
direction, at $z=0$, for droplets with a vortex line along the
$z$ axis and with $N_4=500$ and $N_3=0$, $20$, $50$ and $100$. 
The $^3$He density profiles appear in two disconnected parts
separated by the corresponding $^4$He density profile.
For $N_3=100$, the dashed lines  correspond to a cut at $z=0$
of the densities in Fig.~\protect\ref{fig1}. Lengths are 
in units of {\rm \AA}. }
\label{fig2}
\end{figure}

\begin{figure}
\caption{
Same as in Fig.~\protect\ref{fig1} but with a dopant HCN molecule
in the center of the droplet.
}
\label{fig3}
\end{figure}

\begin{figure}
\caption{From top to bottom panel:
Vortex energy  of the $^4$He$_{500}$+$^3$He$_{N_3}$;
solvation energy of Xe and HCN dopants;
solvation energy of the dopant+vortex complex;
binding energy $|\delta_{\rm X}|$.
The triangles represent results for Xe, the squares for HCN, and
the circles in the top panel are the results for undoped droplets.
The energies are in units of K, and the lines have
been drawn to guide the eye. 
}
\label{fig4}
\end{figure}


\begin{references}
%
\bibitem{Goy92} S. Goyal, D. L. Schutt, and G. Scoles,
Phys. Rev. Lett. {\bf 69}, 933 (1992); 
M. Hartmann, R. E. Miller, J. P. Toennies, and A. F.
Vilesov, Phys. Rev. Lett. {\bf 75}, 1566 (1995); 
K. B. Whaley, Advances in Molecular Vibrations
and Collision Dynamics, Volume 3, 397 (1998);
 J. P. Toennies and A. F. Vilesov, Ann. Rev. Phys.
Chem. {\bf 49}, 1 (1998); 
K. K. Lehmann and G. Scoles, Science {\bf 279}, 2065
(1998).
%
\bibitem{Toe01} J. P. Toennies, A. F. Vilesov, and K. B. Whaley,
Phys. Today, Feb. 2001, 31.
%
\bibitem{bec} M. R. Matthews {\it et al.} Phys. Rev. Lett.
{\bf 83}, 2498 (1999); K. W. Madison {\it et al.}, Phys. Rev.
Lett. {\bf 84}, 806 (2000) and e-print cond-mat/0101051;
B. P. Anderson {\it et al.}, Phys. Rev. Lett. {\bf 86}, 2926
(2001);  J. R. Abo-Shaeer {\it et al.}, Science {\bf 292}, 476 (2001);
S. Inouye {\it et al.}, e-print cond-mat/0104444.
%
\bibitem{Bau95} G. H. Bauer, R. Donnelly, and W. F. Vinen,
J. Low Temp. Phys. {\bf 98}, 47 (1995).
%
\bibitem{Dal00} F. Dalfovo {\it et al.}, Phys. Rev. Lett. {\bf 85},
1028 (2000); M. Pi {\it et al.}, J. Low Temp. Phys. {\bf 121}, 423 (2000).
%
\bibitem{Nau99} K. Nauta and R. E. Miller, Science {\bf 283}, 1895
(1999); {\bf 287}, 293 (2000).
%
\bibitem{Gre98} S. Grebenev, J. P. Toennies, and A. F. Vilesov, Science
{\bf 279}, 2083 (1998).
%
\bibitem{Donnelly}  F. Dalfovo,  Phys. Rev. B {\bf 46}, 5482 (1992);
E. Varoquaux {\it et al.}, Phys. Rev. Lett. {\bf 70}, 2114 (1993);
Y. M. Mukharsky {\it et al.}, Physica B {\bf 194},  591  (1994);
M. Sadd, G. V. Chester and F. Pederiva, Phys. Rev. Lett. {\bf 83},
5310 (1999).  For a discussion about older results see R. J.
Donnelly, {\it Quantized Vortices in Helium II} (Cambridge Univ.
Press, Cambridge, 1991) section 4.7.
%
\bibitem{Bar97} M. Barranco {\it et al.},
Phys. Rev. B {\bf 56},  8997  (1997).
%
\bibitem{Pi99} M. Pi, R. Mayol, and M. Barranco,
Phys. Rev. Lett. {\bf 82}, 3093 (1999).
%
\bibitem{Dup90} J. Dupont-Roc {\it et al.},
J. Low Temp. Phys. {\bf 81}, 31 (1990).
%
\bibitem{T=0} In the density functional $^3$He is treated as a Fermi
fluid in the normal (non superfluid) phase, so that $T=0$
here means that it is low enough to render thermal excitation effects 
negligible in both $^3$He and $^4$He, but it is above
the critical temperature for $^3$He superfluidity.
%
\bibitem{Dal95} F. Dalfovo {\it et al.},
Phys. Rev. B {\bf 52},  1193  (1995).
%
\bibitem{Tan86} K. T. Tang and J. P. Toennies, Z. Phys. D
{\bf 1}, 91 (1986).
%
\bibitem{Atk96} K. M. Atkins and J. M. Hutson,
J. Chem. Phys. {\bf 105},  440  (1996).
%
\bibitem{Jez97} D. M. Jezek {\it et al.}, Phys. Rev. B {\bf 55},  
11092  (1997).
%
\bibitem{note1} It is worth noticing that the displacement of $^4$He
off the vortex axis makes the Feynman-Onsager approximation more reliable
for mixed than for pure $^4$He droplets, since the
possible effects of a non-singular vorticity in the vortex core
[G. Ortiz and D. M. Ceperley, Phys. Rev. Lett {\bf 75}, 4642 (1995);
M. Sadd, G. V. Chester, and L. Reatto, ibid {\bf 79}, 2490 (1997)]
are quenched by the presence of the normal $^3$He component.
%
\bibitem{note2} We want to point out that the four energies involved
in the evaluation of  $\delta_{\rm X}$ are of the order
of $2500$-$3000$ K. This makes its accurate determination
a very demanding numerical task.
%
\end{references}
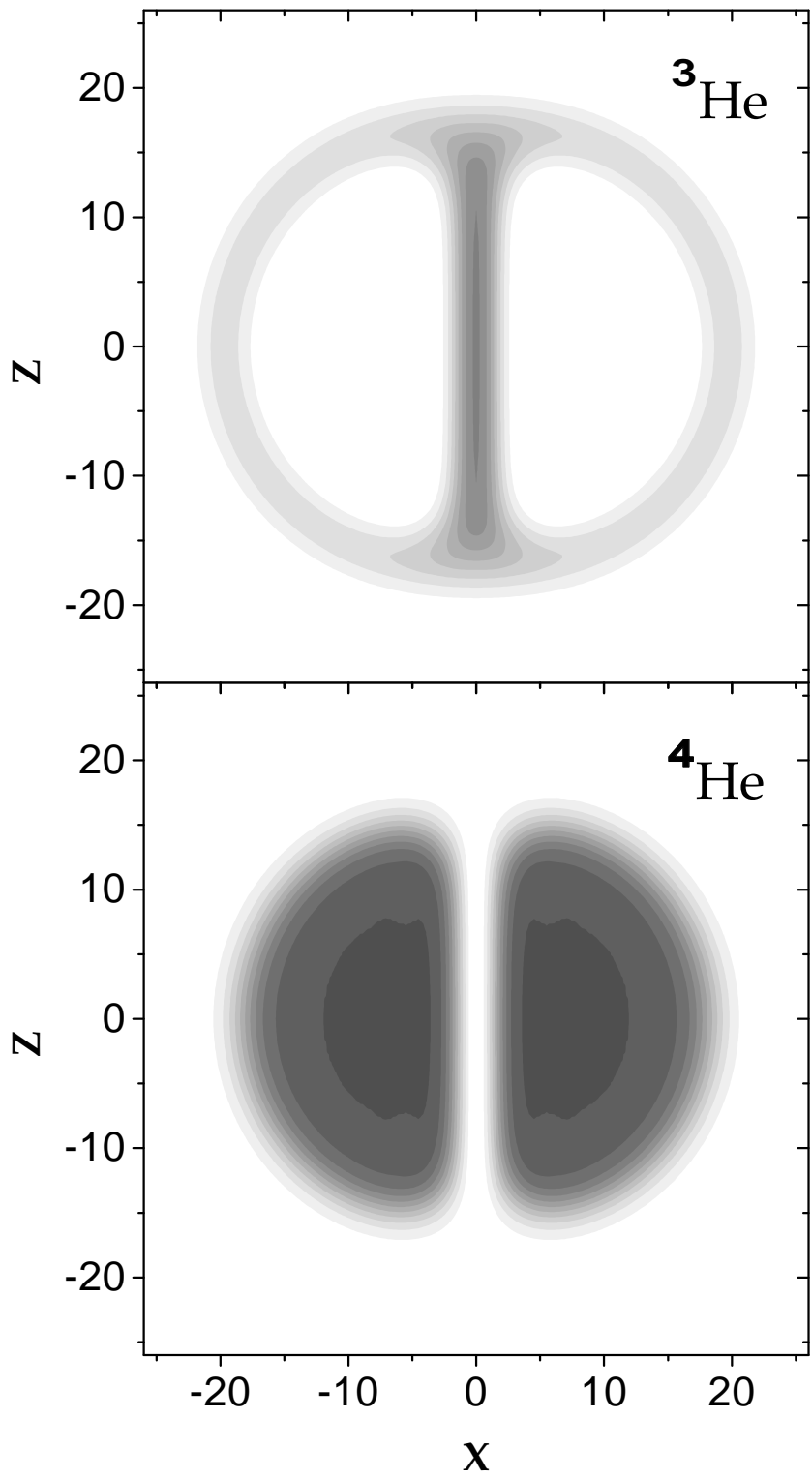
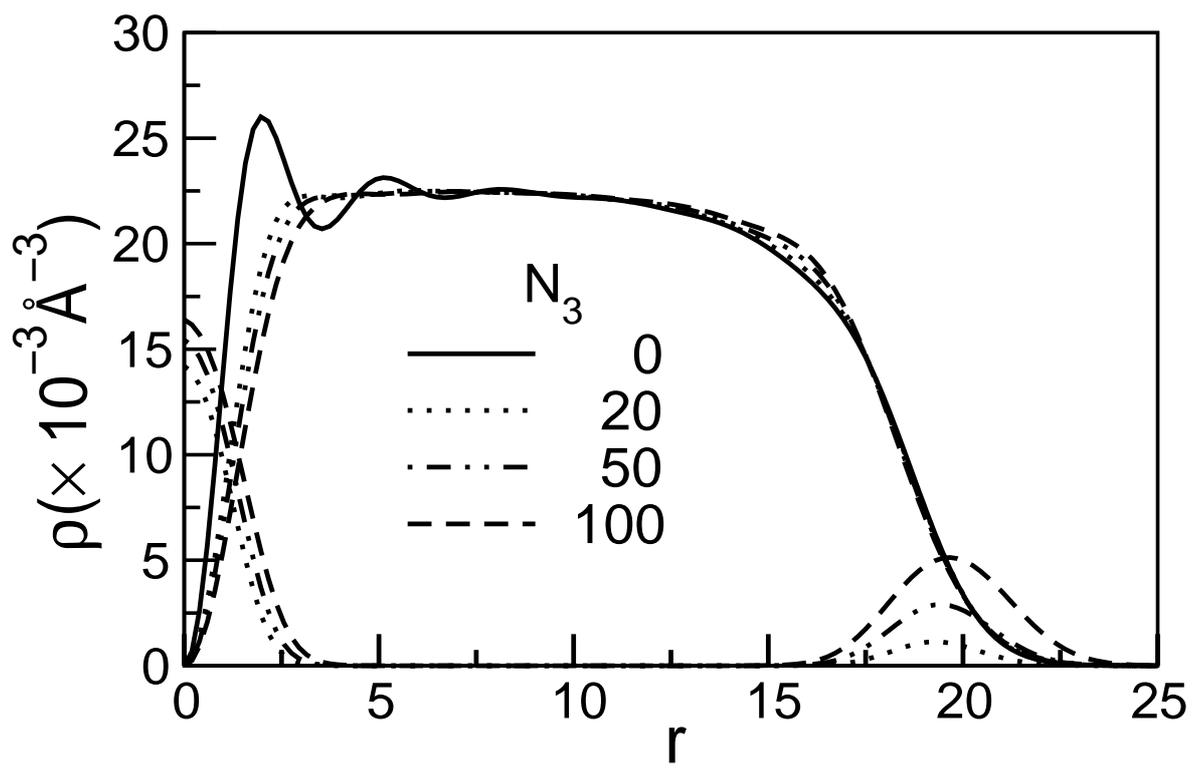
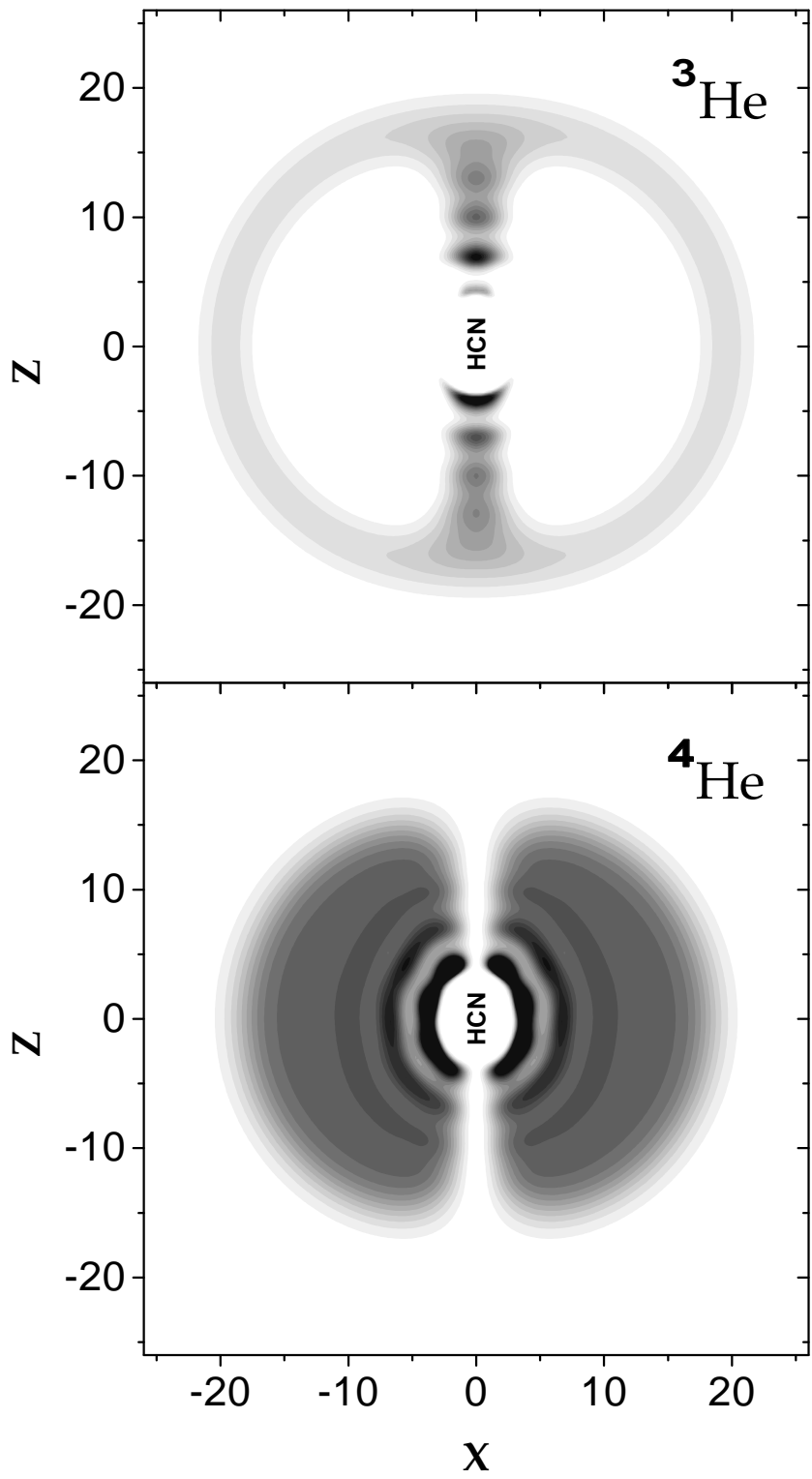
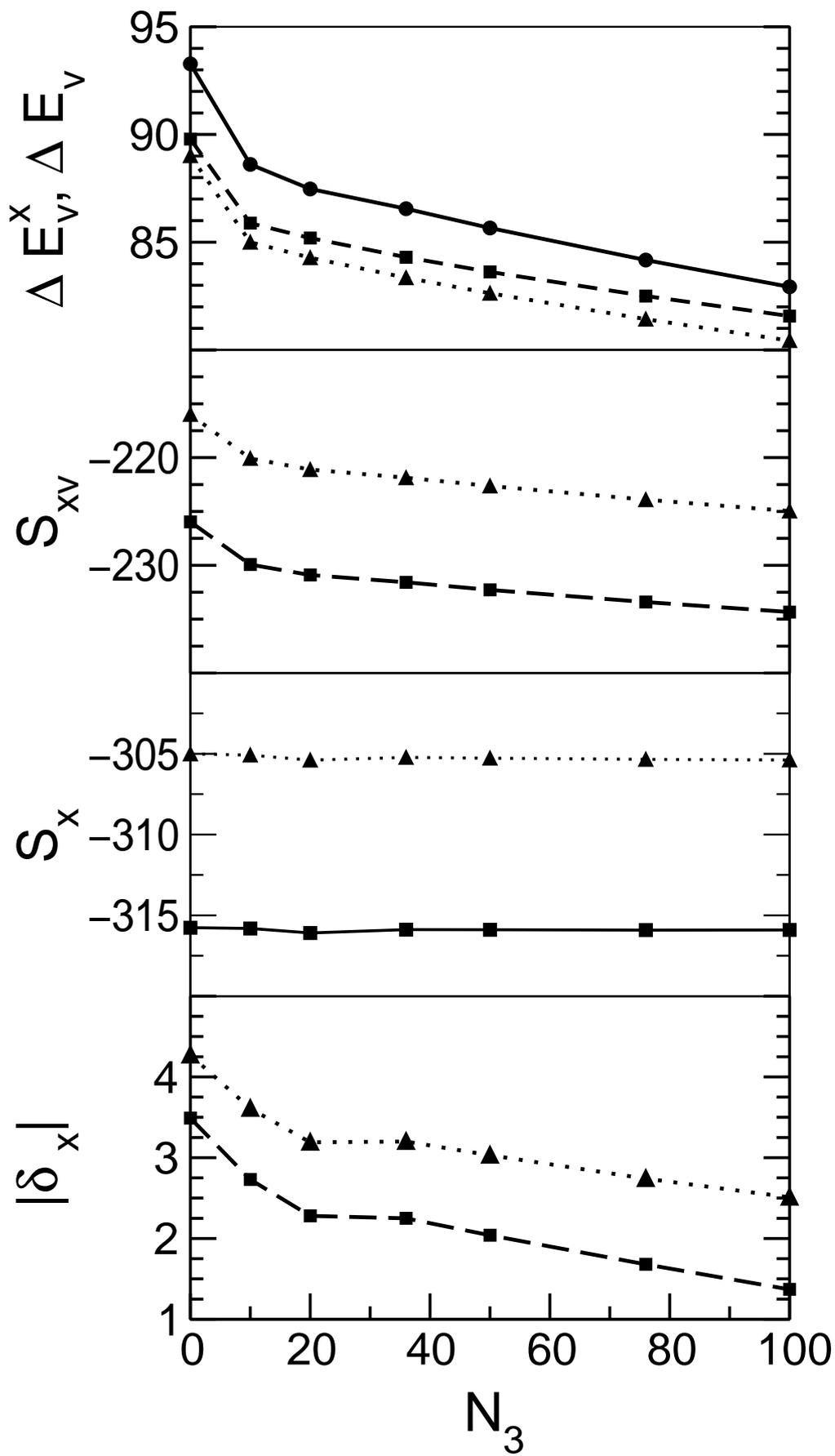
\end{document}